\documentclass[9pt,twocolumn,twoside]{osajnl}
\journal{ol} % Choose journal (ao, aop, josaa, josab, ol)

\setboolean{shortarticle}{true}

\usepackage{SIunits}

\usepackage{color}
\definecolor{mygreen}{rgb}{0,0.5,0} 
\definecolor{myblue}{rgb}{0,0,0.75} 
\definecolor{myyellow}{rgb}{0.87,0.8,0.47} 
\definecolor{mymagenta}{cmyk}{0,1,0,0.12} 
\definecolor{myorange}{rgb}{1,0,0} 
\newcommand{\ctext}[1]{{\color{black}#1}}
\newcommand{\gtext}[1]{{\color{mygreen}#1}}
\renewcommand{\gtext}[1]{{\color{mygreen}} }

\newcommand{\otext}[1]{{\color{black}#1}}

\newcommand{\pumpwl}{\unit{397}{\nano\meter}~}
\newcommand{\sigwl}{\unit{795}{\nano\meter}~}
\newcommand{\NLName}{photo-Kerr effect}

\title{A self-tuning optical resonator}

\author[1,*]{Joanna A. Zieli\'{n}ska}
\author[1,2]{Morgan W. Mitchell}

\affil[1]{ICFO-Institut de Ciencies Fotoniques, The Barcelona Institute of Science and Technology, 08860 Castelldefels (Barcelona), Spain}
\affil[2]{ICREA-Instituci\'{o} Catalana de Recerca i Estudis Avan\c{c}ats, 08015 Barcelona, Spain}
\affil[*]{Corresponding author: joanna.zielinska@icfo.es}

\ociscodes{(190.3270) Kerr effect; (190.1450) Bistability;(190.5330) Photorefractive optics.}

\begin{abstract}
We demonstrate a nonlinear optical resonator that tunes itself onto resonance with an input beam.  In a monolithic Fabry-Perot cavity implemented in rubidium-doped periodically-poled potassium titanyl phosphate, an intensity-dependent refractive index produces line-pulling by multiple free-spectral ranges (FSRs).  In this condition, the cavity passively maintains optical resonance in the face of FSR-scale excursions of the drive laser frequency: when one resonant operating-point becomes unstable, the resonator rapidly transitions to another resonant operating point.  We demonstrate stable second-harmonic generation with no active feedback to laser or cavity.  The self-tuning effect appears to be supported by a very strong, previously unreported optical nonlinearity. 
\end{abstract}

\setboolean{displaycopyright}{true}

\begin{document}

\maketitle

%\section{Introduction}

%\gtext{Dispersive optical nonlinearities underlie a great variety of nonlinear phenomena and optical technologies, including optical memories \cite{ParthenopoulosS1989,YuBook1992}, optical solitons \cite{AgrawalBook2013}, squeezing of light \cite{SchmittPRL1998}, and entangled photon generation \cite{LiPRL2005}. Two broad classes of refractive optical nonlinearities can be identified, fast nonlinearities in which the refractive index depends on the instantaneous intensity, and slow nonlinearities such as the photo-refractive effect \cite{NolteBook2013} in which the  index depends on the history of intensity over a longer time.  Here we observe a new kind of refractive nonlinearity in which the refractive index is linear in the instantaneous field, with a slope that reflects the intensity history of the material. The effect combines  both the strength of slow nonlinearities and speed of fast ones. \ctext{ The strength of the optical bistability caused by this new effect is such that the cavity resonance is broadened to multiple free spectral ranges. What is more, when scanning the laser through the broadened cavity resonance, after the maximum transmission of the peak is reached, it jumps not to the zero like in other systems showing bistability \cite{Hunger10, Sodagar:15}, but directly to the next peak, which we call self-tuning behavior. We demonstrate the utility of this effect with a frequency converter based on the self-tuned resonant cavity.} \mtext{$\lesssim$ 100 words} }

Dispersive optical nonlinearities enable all-optical functionalities including %optical solitons \cite{AgrawalBook2013}, 
optical memories \cite{ParthenopoulosS1989,YuBook1992}, optical switches \cite{YuOE2013}, and optical signal processors \cite{LiuNP2016}. Renewed interest in these devices is driven by their potential to reduce system complexity and latency, for example in chip-level optical interconnects. 

Here we demonstrate a new nonlinear-optical device functionality, the self-tuning optical resonator, applied as a second harmonic generation device.  Many nonlinear optical applications use resonators to boost local intensities, enhancing the nonlinear effect of interest. Higher resonator finesse generally gives greater enhancement, but makes more challenging the task of maintaining optical resonance, which to date requires electronic solutions. This tuning challenge is somewhat mitigated when dispersive  intra-cavity optical nonlinearities can shift the cavity resonance in function of the intra-cavity power, leading to broadened and asymmetric lines without increasing the cavity finesse, hysteresis, and optical bi-stability. The self-tuning resonator employs an extreme version of this effect, in which the line-pulling can reach multiple FSRs, giving the system multiple stable points, all of which are close to  resonance. When the input laser frequency is moved, the system, once the stable point reaches the peak transmission and becomes unstable, automatically finds the next closest to resonance stable point. 

%Stabilization of the cavities that show optical bistability is possible using only proportional feedback \cite{Macke92}, which makes it easier as it does not require demodulation as in Pound-Drever-Hall technique. However, in case of our cavity the bistability at higher powers (above 200mW) becomes so extreme that no proportional feedback is needed to maintain the cavity transmission close to resonance.
%We do not know what causes the effect, KTP has been known to show thermal bistability when used for second harmonic generation to near UV \cite{TorabiGoudarzi2003389}, but this is not thermal as we do not observe change in the effect with the presence of UV light in the cavity.This effect should be investigated further, as optical bistability has a wide range of potential applications. We do not know what causes the effect in terms of material properties, but it has proved really useful to stabilizing our cavity.

%\section{Experimental system}

\begin{figure}[htbp]
\centering
\fbox{\includegraphics[width=\linewidth]{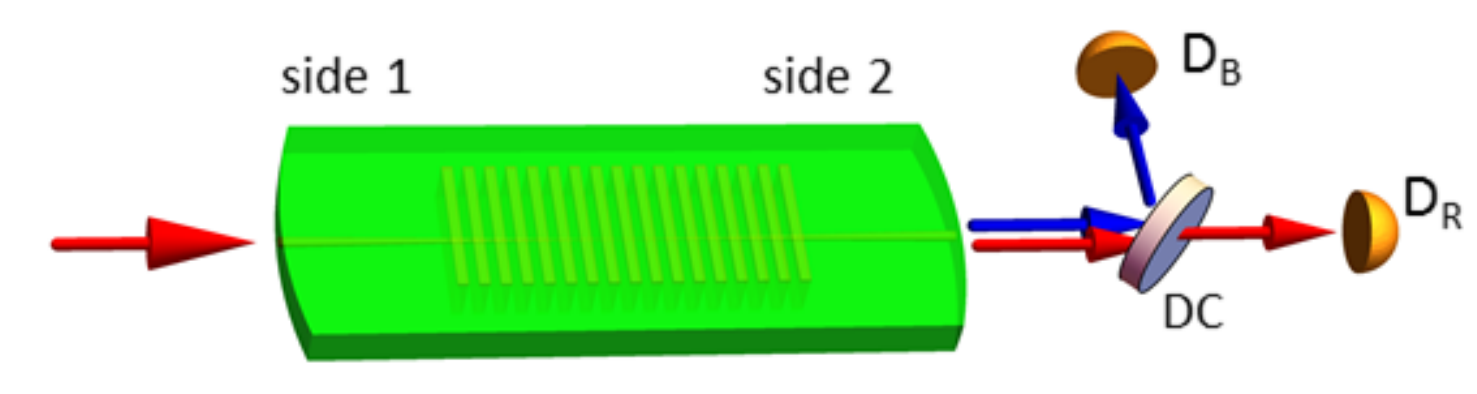}}
\caption{Experimental setup. The abbreviation DC means dichroic mirror, and detectors $D_R$ and $D_B$ collect \sigwl (red) and \pumpwl (blue) light.}
\label{fig:setup}
\end{figure}

The experimental setup is shown in Fig.~\ref{fig:setup}. \ctext{A single crystal device, described in detail in  \cite{Zielinska17}, was constructed from a rubidium-doped potassium titanyl phosphate (RKTP) crystal with dimensions of $\unit{16}{\milli\meter}\times\unit{6}{\milli\meter}\times\unit{1}{\milli\meter}$. The RKTP crystal was poled with poling period \unit{3.16}{\micro\meter} in the middle section (poled volume $\unit{7}{\milli\meter}\times\unit{3.5}{\milli\meter}\times\unit{1}{\milli\meter}$), supplied by KTH Stockholm \cite{Zukauskas:13}. The crystal was spherically polished on both sides with the radius of \unit{10.7}{\milli\meter} and coated to \otext{create desired mode shape and form a stable resonator for} both the fundamental \sigwl and second harmonic (SH) \pumpwl wavelengths. Side 1 is coated to be completely reflective
at \pumpwl, and $84\%$ for \sigwl, whereas side 2 is completely reflective for the \sigwl and $69\%$
reflective for the \pumpwl, giving the resulting finesse 20.5 for the fundamental and 8.4 for the SH. In consequence, when on resonance the fundamental intra-cavity power is approximately 25 times larger than the input power.} The waist of the fundamental beam in the centre of the crystal is $\unit{25}{\micro\meter}$. Independent temperature controls of the central and end sections are used to achieve simultaneous phase matching and double resonance.

%\gtext{The experimental setup, in which we observed the novel Kerr effect is a frequency doubler displayed in Fig.~\ref{fig:setup}, pumped by a \sigwl TA-amplified DBR laser light. The doubly resonant monolithic cavity bases frequency converter is described in detail in \cite{Zielinska17}, the only difference being that in the measurements presented in this work we do not use tuning via the elastooptic effect. The light exiting monolithic cavity at fundamental wavelength \sigwl and second harmonic \pumpwl is split by the dichroic mirror and collected by the two detectors. The monolithic cavity is a doubly resonant device, and the double resonance is achieved by changing the  temperatures of the side sections of the crystal while the center periodically poled section (marked with stripes in Fig. \ref{fig:setup}) is maintained at the phase matching temperature. The cavity has a finesse 20.5 for the \sigwl and 8.4 for the \pumpwl light.}

%\gtext{
%The has been fabricated by our collaborators from KTH Stockholm out of Rb doped KTP crystal, periodically poled with a period of 3.16$\mu$m. The Rb doped KTP crystals are known to be better for periodic poling and less susceptible for gray tracking \cite{LiljestrandOE2016}.  
%}

\begin{figure}[htbp]
\centering
\fbox{\includegraphics[width=\linewidth]{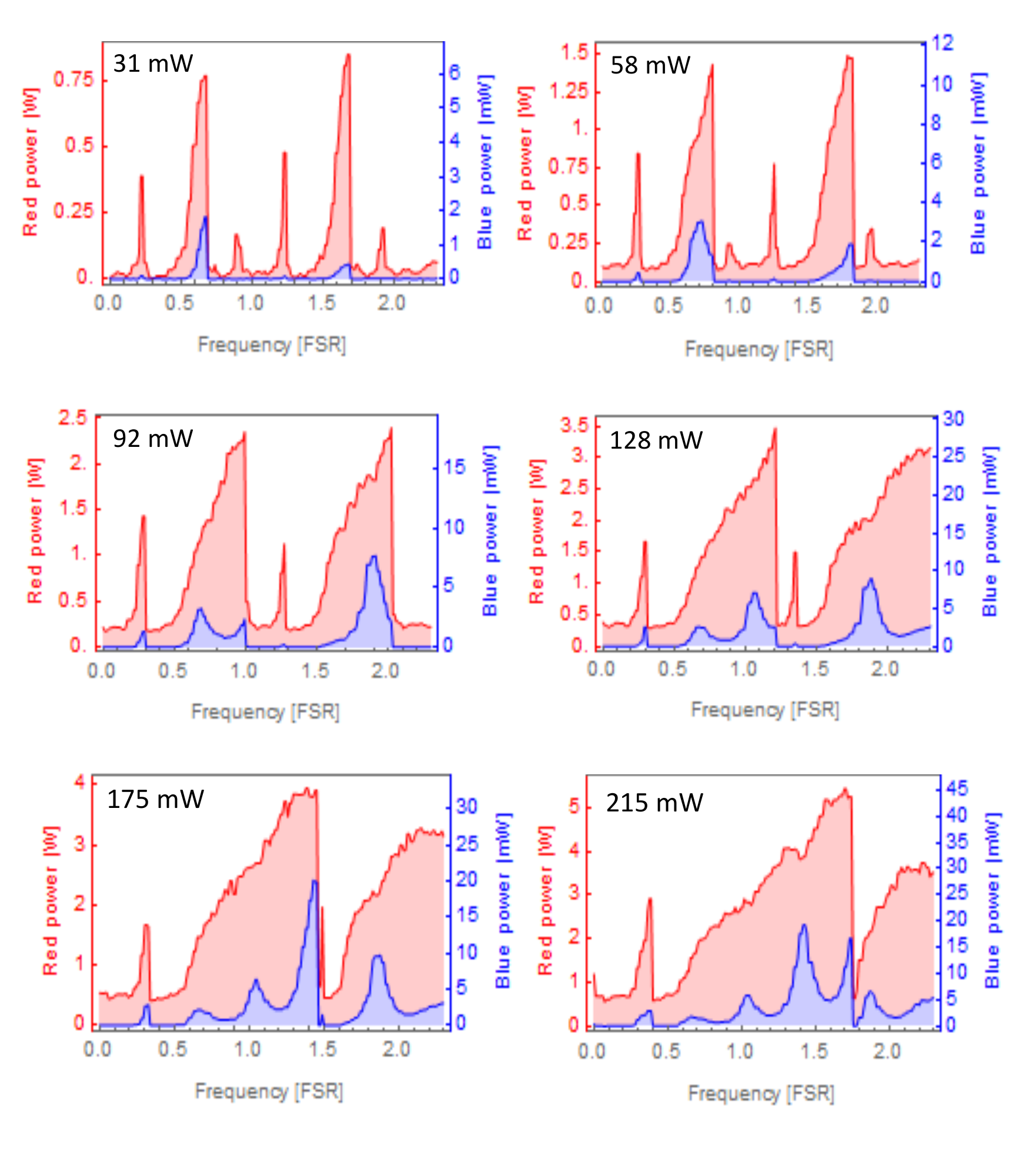}}
\caption{ Red and blue curves show  fundamental and second harmonic intracavity power vs pump laser frequency which is scanned over 2.5 cavity FSR for \sigwl. The scan speed is 10s per FSR, and each graph corresponds to a different pump power as indicated. \otext{Both red and blue intracavity powers are estimated from the respective output power and output coupler transmission values. The red output coupler is calibrated by measuring  output at resonance for known input power and finesse. }}
\label{fig:exppower}
\end{figure}

\begin{figure}[htbp]
	\centering
	\fbox{\includegraphics[width=\linewidth]{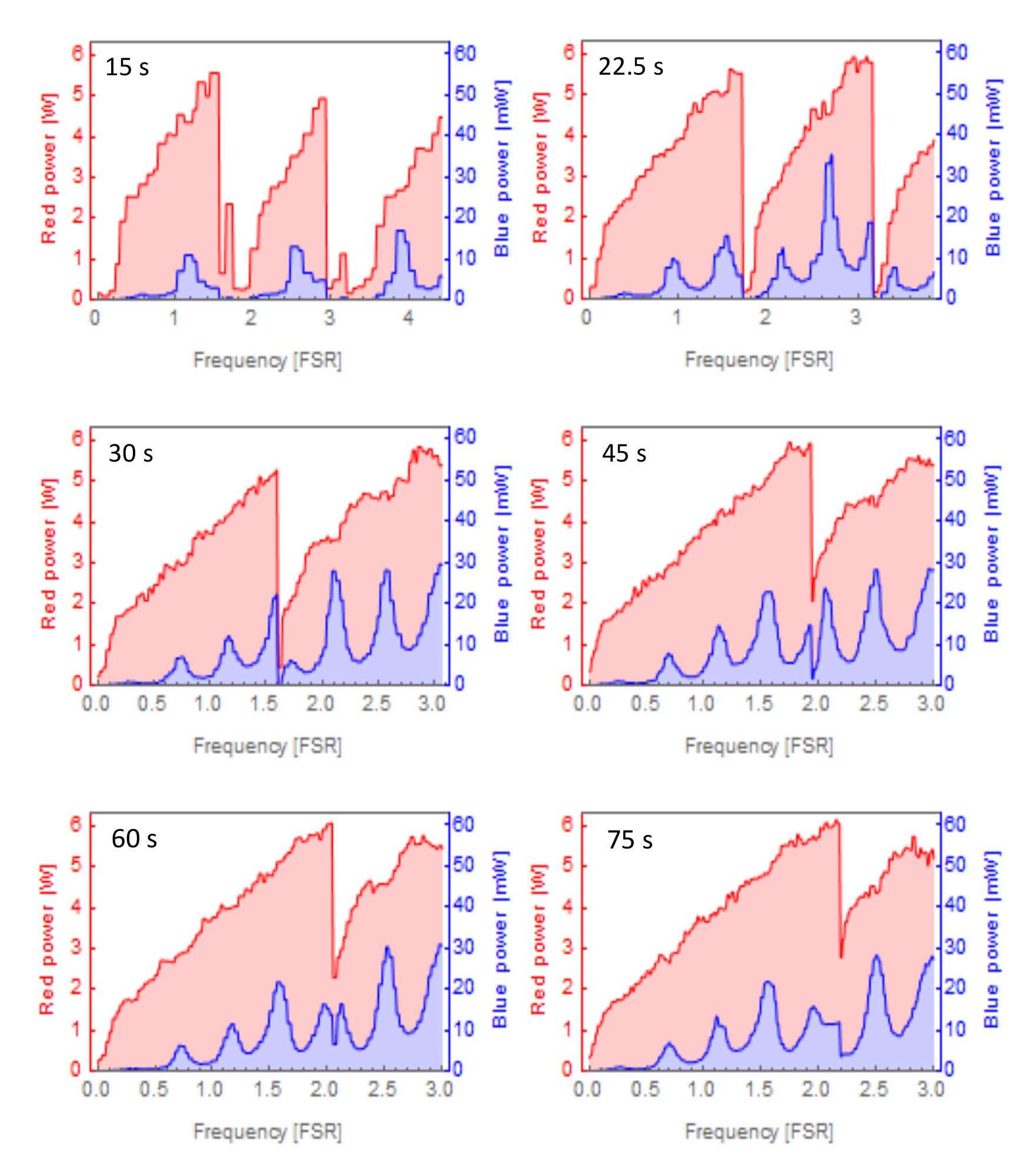}}
	\caption{Red and blue curves show fundamental and second harmonic intracavity power vs pump laser frequency which is scanned over 3 cavity FSR for \sigwl. The pump power is set to 250mW, the graphs correpond to different scan speeds (a total duration of the 3 FSR scan is indicated on each graph).\otext{Both red and blue intracavity powers are estimated from the respective output power and output coupler transmission values.}}
	\label{fig:expspeed}
\end{figure}

The device is pumped with up to \unit{250}{\milli\watt} of tunable  light at \sigwl (\ctext{yielding maximum power density inside the crystal of $\unit{6.3}{\giga\watt/\meter^{2}}$}), and fundamental and SH wavelengths are separately collected at the output. \ctext{The light source used in the experiment was a distributed Bragg reflector laser (Photodigm PH795DBR) emitting cw light with a bandwidth of \unit{1}{\mega\hertz}, amplified by a tapered amplifier (Eagleyard Photonics). The \sigwl light frequency scan was realized by varying the laser current in steps corresponding to \unit{1.5}{\mega\hertz} frequency change. Both fundamental and SH powers at the cavity output were measured using  Thorlabs DET36A detectors (bandwidth \unit{25}{\mega\hertz})}. As shown in Fig. \ref{fig:exppower}, optical bistability is clearly seen when scanning the pump frequency, evidenced by asymmetrical deformation of the cavity resonance, hysteresis, and abrupt transitions from high- to low-power stable points, with all these effects growing with pump power. Kerr bistability is well known to produce such effects, but what we observe here cannot be explained with a simple Kerr nonlinearity.  As shown in Fig.~\ref{fig:expspeed}, the observed effects depend strongly on the speed of the scan, continuing to increase in prominence on long time-scales: We can deduce from Fig.~\ref{fig:expspeed} that the timescale of the {resonance-dragging} effect is $\sim\unit{10}{\second}$, much longer than any optical time-scale in the system, \ctext{such as cavity ring-down time}. At the same time, the jumps from a stable point at one peak to directly to a stable point at another, visible in Fig.~\ref{fig:expspeed}, appear to be very fast (\ctext{compared to the detector response time below \unit{1}{\micro\second}}), suggesting the deformation of cavity spectrum depends on the instantaneous power in the cavity as well as its long-time average, indicating a refractive index depending in (at least) second order on the intensity. \ctext{The maximum observed photo-induced refractive index change is of approximately $\delta n=\unit{5 \times 10^{-5}}{}$, which for the power density in the experiment corresponds to the Kerr constant \footnote{\ctext{Here Kerr constant $K$ is defined so that refractive index change due to power density $I$ at wavelength $\lambda$ yields $\delta n=K \lambda I$. Further on, for convenience we use Kerr coefficient $\kappa$, defined so that $\delta n= \kappa |E|^2$, with $E$ being the incident electric field.}} of $K= \unit{2\times10^{-11}}{\meter/ V^2}$.} Further discussion of the nonlinearity is given at the end of this Letter.

%\gtext{The first observation to be made is that the nonlinear effect causing the dragging of the cavity peaks does not effect the second harmonic light. The spectrum of the blue light is determined by combination of pump power in the cavity, blue cavity resonance spectrum and interference between light generated forward and backward in the cavity.
%}

%\gtext{
%We can deduce from Fig.~\ref{fig:expspeed} that the timescale of the  effect {that deforms the cavity peaks is of order of magnitude of} \unit{10}{\second}. Another observation is that the dragging of the peaks cannot be explained by the simple change of the refractive index for the red light. The jumps from one resonance peak to the next visible in the Fig.~\ref{fig:expspeed} indicate that the effect causing the deformation of cavity spectrum depends on instantaneous power in the cavity as well, therefore it should be at least a third order effect.
%}

The observed nonlinearity has an evident benefit in maintaining resonance in the system: due to the strong nonlinearity, at higher powers the cavity line is pulled by more than a FSR.  As a result, when the system jumps from a resonant  stable point, it drops to another stable point that is also nearly resonant.  As seen in Fig.~\ref{fig:expspeed}, at the highest power and slowest scan, the system maintains $>50\%$ of the intra-cavity pump power after such a jump and then rises toward full power. Resonant behavior can thus be maintained even without frequency stabilization.  We refer to this as cavity self-stabilization.  By setting the frequency to achieve a desired power (up to 95$\%$ of the maximum transmission), we have observed that the intra-cavity power is stable \otext{within few percent} over hours, without any active frequency control. \otext{In in practice we can easily stabilize the cavity length for the red at least 97$\%$ of the maximum
power of the transmission with output power fluctuations of less than 1$\%$ and stability of several hours using a simple slow (\unit{100}{\hertz}) side-of-fringe stabilization, provided the temperatures of the three sections of the crystal are stabilized.} This moreover permits slow adjustment of the temperatures of the various crystal sections to achieve blue resonance. 

%\gtext{
%The observed cavity behaviour results in useful properties in higher powers. Stabilization of the frequency converter can be performed by scanning the cavity slowly in order to increase the intracavity power, and stopping at the desired power (up to 95$\%$ of the maximum transmission). The cavity stays stable for hours, without any feedback locking, and permits for slow adjustment of side temperatures in order to achieve blue resonance without losing the lock. This behaviour causes the narrowband cavity resonances to move so that it behaves like a broadband system.}

%\section{Model}

The bistability effects described above clearly show two time-scales, notable in the fast change from one stable state to another, and in the $\sim \unit{10}{\second}$ accumulation time. The effect does not appear to depend on the presence of the SH  - the bistability occurs also when the SH is not generated, either because it is not resonant, the fundamental pump polarization is rotated, or because the poled section is far from the phase-matching temperature \otext{ of $\unit{39}{\degree C}$} (we observe similar behavior for temperatures ranging from \unit{22}{\degree C} to \unit{55}{\degree C}). The variations of the power of the SH light when pump frequency is scanned depicted in Figs. \ref{fig:exppower} and \ref{fig:expspeed} are determined by combination of pump power in the cavity, SH cavity resonance and interference between light generated forward and backward directions in the cavity, and appear not to be affected by the new nonlinear effect we observe in fundamental light.

While a thermal effect involving the temperature of the whole device could in principle operate on the $\sim\unit{10}{\second}$ timescale, this explanation appears implausible: because RKTP is highly transparent at \sigwl, optical heating of the crystal is mostly due to absorption of the SH light:  the roundtrip loss is estimated to be $30\%$. Thermal changes in the cavity due to presence of fundamental light are hardly noticeable on the crystal temperature sensors, which is not the case with SH light. 

Moreover, \otext{ the observed effects cannot be attributed to fast changes in the local temperature, as thermooptical effect caused by them} would be expected to influence the SH resonance condition more than the fundamental \cite{Kato:02}, whereas the effect we see does not seem to deform the shape of the SH resonances at all.  We can also rule out blue light induced infrared absorption (BLIIRA) as an explanation, despite its long (several minutes) timescale \cite{Tjornhammar:15} because an absorptive effect  alone cannot cause the peak deformation we observe: while absorptive nonlinearities can produce bistability, they cannot pull cavity lines many linewidths away from their cold-cavity frequencies. \ctext{Similarly, the cavity behavior we observe cannot be attributed to the photorefractive damage due to multi-photon absorption (gray tracking) desecribed in \cite{tyminski1991photorefractive}.}

The effect most closely resembles photorefractive effects, in which optically-excited carriers become trapped for long times in impurity levels \cite{gunter2007}  and thereby contribute to the linear refractive index in function of the intensity history of the material. Here, in contrast, what appears to depend on the intensity history is the nonlinear refractive index, or optical Kerr coefficient for the \sigwl wavelength.  For this reason, we call this the \textit{photo-Kerr effect}. These same effects were observed in each of three monolithic RKTP resonators we tested, with one of the three showing cavity-pulling by roughly $15\%$ less than the other two at the same input power. 

%\section{Model}
We find the behavior can be reproduced by a model in which a fast Kerr nonlinearity is present, with a Kerr coefficient that grows at a rate proportional to the intra-cavity red power, and decays on a $\sim \unit{10}{\second}$ time-scale.   The model describes the multi-stable and hysteretic response of the fundamental and SH fields in the driven resonator, subject to  nonlinear propagation equations of the form
{
\begin{eqnarray}
\partial_z A_1(z)&=&ic_1 A_2(z)A_1(z)^* e^{ik_2z-2ik_1(z)z} \\
\partial_z A_2(z)&=&ic_2 A_1(z)^2 e^{-ik_2z+2ik_1(z)z}
\label{eq:fields}
\end{eqnarray}
}
\ctext{where $A_1$ and $A_2$ are  slowly-varying envelopes of the fundamental and SH fields, respectively, (characterized by wavevectors $k_1(z)=[n(\omega_1)+\kappa| A_1(z)|^2]\omega_1/c$ and $k_2$) and $c_j$ are nonlinear coupling coefficients. The refractive index of the fundamental light at frequency $\omega_1$ is denoted as $n(\omega_1)$, and $c$ stands for the speed of light. The Kerr coefficient $\kappa$ evolves slowly as}
{
\begin{equation}
\partial_t \kappa =-\Gamma \kappa + f |A_1|^2,
\end{equation}
}
\ctext{
where $\Gamma$ is the decay rate and $f$ is a coupling constant, allowing the model to describe both fast and slow ($1/\Gamma$) time-scales. 

Since the intracavity power is the variable accessible in the experiment, let us substitute $ P_1=\beta |A_1|^2 $ where $\beta$ is a mode-shape dependent constant with the units of $\unit{}{W m^2/V^2}$.  To simulate a scan, as in Figs. \ref{fig:exppower} and \ref{fig:expspeed}, we define a small time step $\tau$ and update $\kappa_j \equiv \kappa(j \tau)$ as 
\begin{equation}
\kappa_{j+1}=M\kappa_j+F P_1^{(j)}
\label{eq:kerr}
\end{equation}
where $M = \exp[-\tau\Gamma]$, $F = \tau f \beta^{^-1}$ and update the fields changing their frequency with the rate according to the scan speed and finding the steady state solution of the Eq. (\ref{eq:fields}).

\begin{figure}[htb]
	\centering
	\fbox{\includegraphics[width=0.95\linewidth]{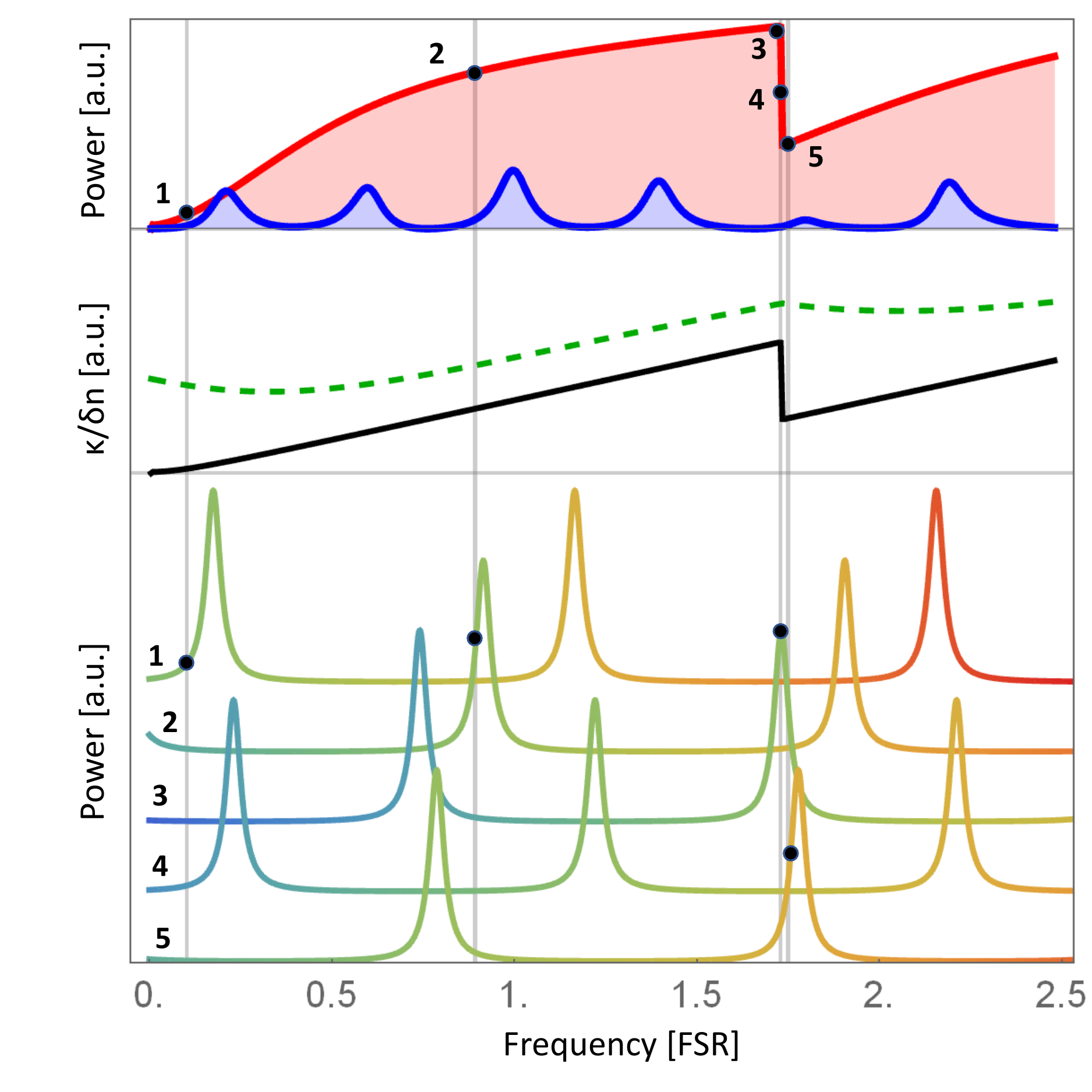}}
	\caption{Model results illustrating self-locking by \NLName. Upper part: Intracavity power of fundamental (red) and SH (blue) wavelengths as the fundamental pump frequency is scanned.  Conditions: scan rate \unit{10}{\second} per FSR, \unit{250}{\milli\watt} input power, active section at the phase matching temperature.  Middle part:  Kerr coefficient $\kappa$ (green dashed curve) and refractive index change $\delta n=\kappa |A_1|^2$ (black curve). The cavity spectrum shift is proportional to $\delta n$. Lower part: cavity transmission for the fundamental at five representative points of the scan (marked in upper part). }  
	\label{fig:theoscan}
\end{figure}

\begin{figure}[htb]
\centering
\fbox{\includegraphics[width=0.9\linewidth]{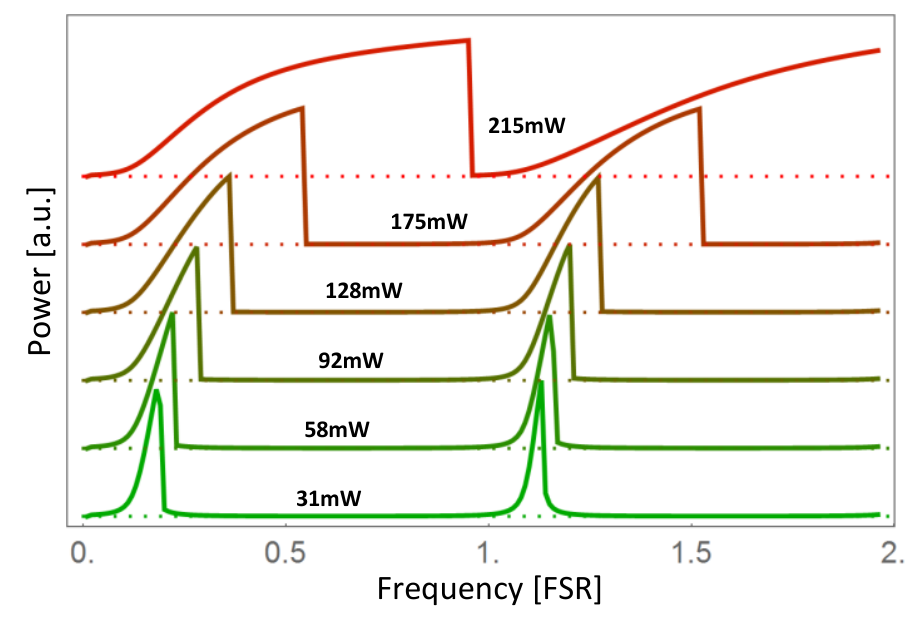}}
\caption{Cavity scans calculated from the model for powers \unit{31}{\milli\watt} (green), \unit{58}{\milli\watt}, \unit{92}{\milli\watt}, \unit{128}{\milli\watt}, \unit{175}{\milli\watt}, \unit{215}{\milli\watt} and \unit{250}{\milli\watt} (red) and scan duration of \unit{30}{\second}. {Curves are offset with baselines indicated by dotted lines.}}
\label{fig:theopwr}
\end{figure}

Results from the model are shown in Figs.~\ref{fig:theoscan}, \ref{fig:theopwr} and \ref{fig:theospeed}.  Fig.~\ref{fig:theoscan} illustrates the bistability mechanism and shows how the \NLName~produces self-stabilization near a cavity resonance, {presenting results  calculated at the phase-matching temperature.} Following the fundamental light frequency scan, at point 1 (marked in the upper part of the figure and), the red field encounters a resonance. At point 2, the side-of-resonance condition is maintained by the combined effects of resonance and nonlinear refractive index. $\kappa$ accumulates, allowing the resonance to shift by more than one FSR.  When the top of the resonance peak is reached (point 3), the resonance cannot shift anymore and the system becomes unstable. The power starts to drop and cavity resonance quickly retreats (point 4). Because the peak was shifted by more than one FSR, the system encounters another stable side-of-resonance condition (orange in the figure) as the nonlinear shift reduces (point five). As the scan proceeds the resonances once again shift and the process repeats.

The dependence of cavity peak deformation on the power calculated from the theoretical model is presented in Fig.~ \ref{fig:theopwr}. For simplicity, the calculation is performed out of phase matching temperature and only the fundamental light power is shown {in Figs. \ref{fig:theopwr} and \ref{fig:theospeed}}. Similar dependence on the scan speed can be found in the Fig.~\ref{fig:theospeed}. The model reproduces qualitatively the behaviour of the cavity. 

\begin{figure}[H]
\centering
\fbox{\includegraphics[width=0.9\linewidth]{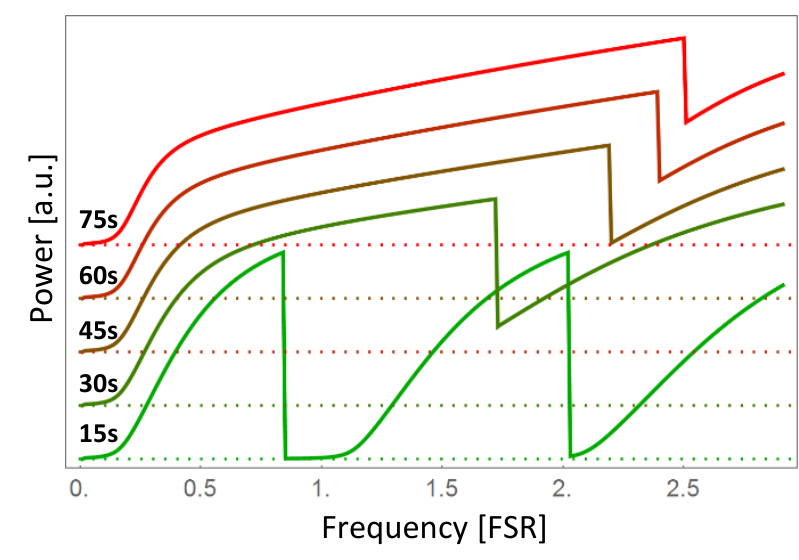}}
\caption{Cavity scans calculated from the model for scan durations 15s (green), 30s, 45s, 60s and 75s (red) for pump input power of 250mW. {Curves are offset with baselines indicated by dotted lines.}}
\label{fig:theospeed}
\end{figure}

Model parameters that give reasonable agreement with observation are decay time $\Gamma^{-1}=\unit{15.13}{s}$ (corresponding to $M=0.9967$), $\beta^{-1}F=1.53\times 10^{-8}\unit{}{W^{-2}}$, the initial Kerr coefficient multiplied by the inverse of geometry constant $\beta^{-1} \kappa_{0}=4 \times 10^{-6} \unit{}{W^{-1}}$, and time step  $\tau = \unit{50}{ms}$. The same $M$, $F$, $\kappa_{0}$ and $\tau$ parameters are used for results in Figs.~\ref{fig:theoscan}, \ref{fig:theopwr} and \ref{fig:theospeed}.}

%\section{Conclusion}

%\gtext{We describe a previously unreported optical nonlinearity, in which the optical Kerr coefficient of a material strongly depends on the long-time average of the intensity in the material. The effect is clearly observed in a RKTP monolithic optical resonator, through dispersive optical bistability features that depend on long-time average of intra-cavity power. Modeling with nonlinear optical propagation equations well reproduces the observed behavior, and indicates that the new effect is far stronger than the ordinary Kerr effect in this scenario.  At moderate input powers, the effect is sufficiently strong as to produce cavity self-locking, i.e., cavity mode-pulling maintains the cavity near peak resonance  even as the laser frequency changes by more than a FSR, greatly simplifying the stabilization of the cavity used as a frequency converter.  }

We have described a self-tuning nonlinear optical resonator, in which a strong dispersive optical nonlinearity produces cavity line-pulling that can reach up to multiple FSRs. In these conditions, optical multi-stability maintains resonance of the cavity, because multiple stable points involve significant intra-cavity power and near-resonance.  We have used the self-tuning device  for cavity-enhanced second-harmonic generation with no active tuning. The effect appears to derive from a previously unreported optical nonlinearity, in which the optical Kerr coefficient of a material strongly depends on the long-time average of the intensity in the material. Modeling with nonlinear optical propagation equations well reproduces the observed behavior, and indicates that the new nonlinearity is far stronger than the ordinary Kerr effect in the host material.    

%\mtext{to be revised}
%We observed a Kerr effect when scanning the wavelength of the light injected into a monolithic frequency doubling cavity, with a particular feature of deformation of the cavity peak dependent on the speed of wavelength scan. We presented a qualitative model of this behaviour as a cumulative Kerr effect, where the Kerr coefficient depends on how much power has been in the cavity before, displaying a memory-like characteristics with timescale of 10s. The effect is very useful for the cavity stabilization purposes as it causes an auto-tuning behaviour, making the cavity adjust its resonance when the wavelength moves of the input light changes, resulting in being close to resonance always provided that the power is sufficient.

\section*{Funding Information}

This work was supported by European Research Council (ERC) projects AQUMET (280169) and ERIDIAN (713682); European Union QUIC (641122); Ministerio de Economia y Competitividad (MINECO) Severo Ochoa programme (SEV-2015-0522) and projects MAQRO (Ref. FIS2015-68039-P), XPLICA  (FIS2014-62181-EXP); Ag\`{e}ncia de Gesti\'{o} d'Ajuts Universitaris i de Recerca (AGAUR) project (2014-SGR-1295); 
Fundaci\'{o} Privada CELLEX; J.Z. was supported by the FI-DGR PhD-fellowship program of the Generalitat of Catalonia.

\section*{Acknowledgments}

We thank Carlota Canalias and Andrius Zukauskas from KTH Royal Institute of Technology in Stockholm for valuable discussions and for fabricating the RKTP crystals. 

\bibliography{kerr}

%\bibliographyfullrefs{kerr}
%\section*{References}

\end{document}